# Distributed Query Processing Plans generation using Teacher Learner Based Optimization

Vikash Mishra[a], Vikram Singh[a]

[a] Department of Computer Engineering, National Institute of Technology, Kurukshetra, Haryana 136119, India,
Contact: mishravikash03@gmail.com, viks@nitkkr.ac.in

With the growing popularity, the number of data sources and the amount of data has been growing very fast in recent years. The distribution of operational data on disperse data sources impose a challenge on processing user queries. In such database systems, the database relations required by a query to answer may be stored at multiple sites. This leads to an exponential increase in the number of possible equivalent or alternatives of a user query. Though it is not computationally reasonable to explore exhaustively all possible query plans in a large search space, thus a strategy is requisite to produce optimal query plans in distributed database systems. The query plan with most cost-effective option for query processing is measured necessary and must be generated for a given query. This paper attempts to generate such optimal query plans using a parameter less optimization technique 'Teaching-Learner Based Optimization' (TLBO). The TLBO algorithm was experiential to go one better than the other optimization algorithms for the multi-objective unconstrained and constrained benchmark problems. Experimental comparisons of TLBO based optimal plan generation with the multi-objective genetic algorithm based distributed query plan generation algorithm shows that for higher number of relations, the TLBO based algorithm is able to generate comparatively better quality Top-K query plans.

## 1. INTRODUCTION

Database systems of different type use various techniques to identify optimal query plans. In many application domains, end-users are more interested in the most important (top-k) query answers in potentially huge answer space [1]. A distributed database encompasses coherent data, spread across various operational autonomous sites of a computer network [3]. With the increase in the number of users or expanding organization requirement, the size of database networks also expands. In recent years, it is observed that, in any distributed database the number of data sources and the amount of data has been growing very fast to meet the challenge of growing popularity or business expansion. A Distributed Database Management System (DDBMS) deals with managing such distributed databases. DDBMS provides access to user via a simple and unified interface over disperse databases through different transparency mechanism, due to which a user feel as if they were not distributed [14]. The query processing is a prime activity in database and it is controlled by DBMS. The performance of a DBMS is determined by its ability to process queries in an effective and efficient manner [15]. The query processing in distributed databases is more intricate, as there various parameters or constraints involve and they are affecting performance [1].

Query processing connects to many database research areas, including query optimization, indexing methods, and query languages [6], [7]. As a consequence, the impact of efficient processing of query is becoming apparent in an increasing number of applications. One common way to identify the top-k objects is scoring all database objects on some scoring function. An object score acts as a valuation for that object according to its characteristics (e.g., price and size of house objects in a real estate database, or color and texture of images in a multimedia database). Query plans are usually evaluated by multiple scoring predicates or objective functions that contribute to the total object score. A scoring function is therefore usually defined as an aggregation over objectives or scores [19].

In DDB, data is distributed over the multiple autonomous logical sites. The data distribution policy decides the manner of distribution of logical units of operational data. There are two alternative of distribution of data in distributed database: full replication or partition etc. due to which a given relation can be found in more than one sites [4]. This distribution of data imposed challenges to query processing, as for a user query there are multiple semantically valid query equivalent plans (QEPs) possible. These alternatives are equivalent in terms of outcome as they retrieve same set of database objects or records. Thus selecting best alternatives for processing from generated pool is a decisive task to query optimizers. In DBMS, query optimizer is an essential component, with primary objective is to choose optimal (best) solution [3]. In optimization, an alternative is considered better or fitter than other based on the objective function values for the query result generation. In DDB systems, query processing and optimization is constrained and subjective by different cost parameters such as, communication cost, local processing cost, optimization cost, query localization cost etc.

For a user query, first step involve the identification of the relevant logical units of query and the data relevant to query is usually available at different sites. The query processing, thus, would involve transmission of data between these sites. These data transmissions, along with local data processing, constitute a distributed query processing (DQP) strategy for a user query [6]. In DQP, the distributed query is parsed before arriving at an effective query processing strategy for it [12]. This strategy comprises of effective and efficient query processing plans that would decompose the distributed queries into local sub-queries to be executed at their respective sites. Also, the logical order of relational operator and the site (control site) at which the results of the sub-queries are integrated is also part of this plan. The finally integrated result is provided as the answer of the query. Thus the DQP strategy aims to generate query processing plans that reduce the amount of data transfer between participant sites by selecting the appropriate copy of data for query result retrieval and thereby reduces the overall distributed query response time [14],[11]. This paper focuses on generating optimal query processing plans for distributed relational queries.

In the proposed heuristic, optimality on query plans is based on the function of different cost models or function; cost models are assumed attributes of query processing. Each of the query equivalent plans has associated set of pre-computed cost values, based on which optimization is performed. Computation of cost is according to the different primitives proposed in next section. In many real-world problems, optimization is based on two or more objective functions simultaneously. These problems are known as multi-objective optimization problems (MOPs), and solution involves finding not one optimal solution, but determine a set of solutions that represent the best possible trade-offs among the objective functions being optimized. Such trade-offs constitute the Pareto optimal set and their corresponding objective function values form the Pareto front for a user query in distributed database [18].

Optimization related computation in most of the evolutionary and swarm intelligence-based algorithms are probabilistic. This is mainly involved controlling common parameters, such as search space size, number of generations, elite size, etc. In addition computation of common control parameters, algorithm-specific control-parameters evaluation is also required, such as in GA rate of mutation and crossover rate, similarly, inertia weight and social parameters in PSO. The proper regulation of algorithm specific-parameters is a very critical aspect, as it affects the overall performance of the optimization algorithms. The improper regulation of algorithm-specific parameters may lead to an increased computational effort or yields a localized optimal solution. Hence there is need of an optimization approach, in which tuning of algorithm specific parameters can minimized or eliminated in the task of optimization. Recently Rao et .al, introduced the Teaching Learning Based Optimization (TLBO) algorithm, which requires only the common control parameters and does not require any algorithm-specific control parameters [21],[22],[23]. Other evolutionary algorithms require the control of common control parameters as well as the control of algorithm-specific parameters. The burden of regulation on control parameters is comparatively less in the TLBO, thus the TLBO is simple, effective and involves less computational effort. There is another advantage of using TLBO, that it uses some well established benchmark functions to evaluate the final fitness of an alternative. Hence, in the present work, TLBO is used on multi-objective query optimization on unconstrained test functions, and performance is compared with other nature-inspired optimization algorithms such as Vector Evaluated Genetic Algorithm (VEGA) and Aggregation based Genetic Algorithm.

## 1.1. Related Work

Query Optimization in distributed database systems is NP-complete problem as for given user query multiple semantic equivalent plans [5]. In [6], shows that processing all generated query plans leads to a problem an exhaustive search and it not computationally viable. Further, this being a combinatorial optimization problem [9], it can be addressed by various optimization techniques based on heuristics like greedy, evolutionary, and randomized [6],[4],[5],[8],[16],[17]. However, efficiency of these techniques is affected by the unconventional behaviour, in specific instances, of the problem [17]. In [18], an approach that generates "close" query plans with respect to the number of sites involved and the concentration of relations in the sites for a distributed relational query is given. As per [18],[19], query processing over lesser number of sites would be more efficient and thus query plans involving fewer sites need to be generated. Such query plans, referred to as "close" query plans, are generated using the genetic algorithm (GA) [18] [19], without considering the communication and local processing cost on optimization of QEPs. None of the existing approach considered the fundamental cost models for optimization. In [27],[28], [29] optimal query plans are generated according to various customized cost models, in this paper we have accommodated the localization cost as integral part of query communication cost and subsequently local processing cost based on predicate selectivity of local operator. A optimization algorithm's performance is entirely based on algorithmic parameters [23], in this paper we have applied TLBO, which not required tuning of any algorithm specific parametric during generating of optimal query plans.

## 1.2. Contribution and Outline

The primary contribution of the current work is to exhibit the use of parameter-less optimization over the genetic algorithm inspired for query optimization for distributed query processing. The overall performance of TLBO based approach is better as it is independent from a tuning of any algorithm specific parameter during optimization. The entire optimization cost is eliminated in TLBO, while GA based optimization it is an additional cost component. Another contribution of this paper is to analysis, the effect of various proposed cost model/functions for DQP.

Section 2, discussed fundamental of distributed query processing and describe the proposed design objectives for 'Optimal Query' generation. Example is also drawn to demonstrate for the proposed heuristics. In section 3, fundamental of TLBO are discussed and algorithm of the same for query optimization is presented. Section 5, experimental results of performance of Aggregation based GA, VEGA and TLBO on 'Optimal Top-K Query' generation are shown via various graphs.

## 2. DISTRIBUTED QUERY PROCESSING

Query processing in distributed database involves lot of data communication between participant sites. This inter-site communication/transfer of relevant data is dominant factor or cost to constraint the overall query processing and optimization. The degree of data transfer is directly related with the heterogeneity in sites accessed in a query plan. For instance, in a query plan all the relations are accessed from different sites, then this strategy will lead to highest amount of data transfer among sites and thus it is not preferred by a query optimizer. In DQP, the various costs incurred are CPU, I/O and the site-to-site communication cost. Among these, the inter-site communication cost is the dominant cost. In order to process a user query in distributed database system, the data required may have to be obtained from several sites distributed over a computer network. Furthermore, as the number of sites containing the relations accessed by the query increase, the number of possible valid query plans also increases. So it becomes imperative to arrive at a query processing plan that entails an optimal cost for query processing. However, the number of such possible query plans increases exponentially with increase in the number of relations in the query and also with increase in the number of sites containing them [12]. Thus, a large search space comprising all possible query plans needs to be explored in order to compute the optimal query plans.

Query processing in such environment is difficult task for query processor, as for a user query there are multiple query equivalent alternatives possible. These query alternatives are equivalent on the output terms. The semantics of the all equivalent are similar as they access similar set of relations but from different operational sites. The database allocation policy plays an important role on the selection of appropriate sites of relations. Selecting best alternatives for processing from the generated pool of alternatives is a critical task to query optimizers, whose primary objective is to choose optimal (best) solutions [3]. The selection of optimal plans is according to trade-offs among the various design objectives or heuristics. In this paper we propose cost model or deign objectives for query processing based on which the optimality on QEPs are applied. In next sub section, the various proposed cost model are discussed.

## 2.1. Heuristic's of Design Objective

*Query Affinity Cost (QAC):* This is first design objective; it indicates the degree of heterogeneity in a QEP on the number of sites accessed during result generation for given user query. For a given user query, a query equivalent plan that involves less or may be least number of sites is measured better than the other alternatives. If a QEP is accessing similar sites for relevant relation for result generation of a user query it is better than a QEP in which more sites are accessed for the same set of relations. In case more than one such plans are present search space, then a plan having sites with higher concentration of same relations is considered better. The formulation for the QAC evaluation is according to below given equation for a query plans:

$$QAC = \sum_{i=1}^{M} \frac{K_i}{N}\left(1 - \frac{K_i}{N}\right) \quad (1)$$

In the equation, M is the total number of sites required in the query plan, N is the total number of relations accessed by query plan and $K_i$ is the number of times the ith site is accessed by query plan.

*Query Localization Cost (QLC)*: The 'Query Localization' indicates two design objectives, first it quantify the degree of communication between two different sites in a query plans (QP) and second, it plays crucial role on deciding the control site for query answering of respective QP. In the existing cost models by various distributed query processing approaches, the cost of deciding the control site is never considered. We have emphasizes the importance of the same and proposed QLC design objective by incorporating the importance of localization. The control sites for query plan execution is decided based on the size of relations stored in participating site and communication cost is evaluated purely based on the tuples or db records selected during local processing. A QEP is mapped in a query graph on which sites as nodes and edges as join selectivity (QLC values) are mapped. Computing join selectivity for a QP on dynamic and is distributed environment is biggest challenge. QLC between two sites, $S_1$ to $S_2$ represents the ratio of size of relation at that site and sum of sizes of total relations in FROM clause of query. For a given query plan the minimum QLC can is computed as follows,

$$QP_{Cost} = MIN_{i=1}^{Nr}\left[\sum_{j=1 \&\& j \neq i}^{Nr} \frac{Size(R_{S_j})}{\sum_{k=1}^{Nr} Size(R_k)}\right] \quad (2)$$

Where, query localization or communication cost is indicated by $QP_{cost}$ of a QEP, function MIN to evaluate the minimum value for i=(1 to $N_r$), total number of relations accessed by the user query or number of relations in the FROM clause of user query is $N_r$, Size($R_{sj}$) is total number of data tuples in a relation present at site $S_j$, Size($R_k$) is number of tuples in relation $R_k$. QLC between self communication is zero, eg. QLC among $S_i$ and $S_i$ will be zero. QLC is a important design objective as in distributed database system, communication cost is dominant cost component.

*Local Processing Cost (LPC)*: Final design objective is LPC, which is a primarily concern by database relations stored on a local site and selectivity of database operators. There are various relational operators such as Selection, Projection etc., quantify the value of LPC of a QEP. The operator selectivity is the measure of LPC. In other words, usually LPC is dependent on number of memory accesses or memory fetch for transferring set of tuples from secondary memory to main memory, while retrieving data for local relations. We have categorized two component of LPC, first is due to local processing computation on relations at remote sites and second due to local processing computation on control site (for final result preparation). The second component LPC is important as, final result is integrated on the control sites a supplied to the user in desired structure. For evaluation following are equation are used,

Relation Processing Cost-

$$RPC = N_t * S_r / \sum_{k=1}^{Nr} N_t(k) \quad (3)$$

(a) LPC for Remote Site used in Query Plan
$$RLPC = Max_{i=1 \ to \ Rs}[RPC(i)] \quad (3.a)$$

(b) LPC for Control Site used in Query Plan
$$CLPC = Max_{i,j=1toNr}[N_t(JOIN(R_i,R_j)) * S_j((JOIN(R_i,R_j))/\sum_{k=1}^{Nr} N_t(k)] \quad (3.b)$$

LPC of a query plan QPi
$$= \sum_{k=1}^{Sqp-1}(RLPC(k)) + (CLPC - Max_{i=1toNr \&\& R(i)CS\&JOIN}[RPC(i)]) \quad (3.a+3.b)$$

Where total number of tuples is $N_t$ and total number of relation in user query is $N_r$, $S_{qp}$ represents the total number of sites accessed by user query, $R_s$ is total number of relations stored in local site, $S_r$ is selectivity measure of relation R on local site, $S_j$ is selectivity measure of Join relational operator and $N_j$ is number of Joins operation for a query plan.

**Example**: In distributed database systems relations are spread across multiple sites, and multiple copies of a relation stored/ maintained in different sites. A working scenario is shown in table 1 (a), which depicts a typical DDBS allocation of 8 database relations among 16 sites. Relation-Site matrix gives details about allocation of relation and sites. eg. Relation $R_1$ is stored in $S_1$, $S_2$, $S_3$, $S_5$, $S_6$, $S_8$, $S_{10}$ and it is assumed that $R_1$ is replicated entirely not in fragmented or partitioned form. In some database the fragmented or partitioned copy is replicated in various sites. The objective of keeping multiple instances of database relation in multiples sites is to achieve higher reliability and availability. Distributed query processing also tries to select the proximate copy of relation to the user. So, once user or application pose a query on the DDBS, it is critical to identify the relevant sites of a relation. The first step in the query retrieval of results and based on the RSM a query optimizer identify the sites on which particular relation store.

Table 1
(a) Allocation schema (Relation-Site Matrix (RSM)) (b) subset of Query Plans (QP's) of query $Q_1$

| Site / Relation | $R_1$ | $R_2$ | $R_3$ | $R_4$ | $R_5$ | $R_6$ | $R_7$ | $R_8$ |
|---|---|---|---|---|---|---|---|---|
| $S_1$ | 1 | 1 | 1 | 1 | 1 | 0 | 0 | 0 |
| $S_2$ | 1 | 1 | 1 | 1 | 1 | 0 | 0 | 0 |
| $S_3$ | 1 | 0 | 0 | 0 | 0 | 1 | 0 | 1 |
| $S_4$ | 0 | 0 | 0 | 0 | 0 | 1 | 0 | 0 |
| $S_5$ | 1 | 1 | 1 | 0 | 0 | 1 | 1 | 0 |
| $S_6$ | 1 | 0 | 0 | 0 | 0 | 1 | 1 | 0 |
| $S_7$ | 0 | 1 | 1 | 1 | 1 | 1 | 0 | 0 |
| $S_8$ | 1 | 1 | 1 | 1 | 0 | 0 | 1 | 1 |
| $S_9$ | 0 | 0 | 0 | 0 | 0 | 0 | 0 | 1 |
| $S_{10}$ | 1 | 0 | 1 | 0 | 0 | 0 | 0 | 0 |
| $S_{11}$ | 0 | 1 | 1 | 1 | 0 | 0 | 1 | 0 |
| $S_{12}$ | 0 | 1 | 0 | 0 | 0 | 0 | 0 | 1 |
| $S_{13}$ | 0 | 0 | 1 | 0 | 0 | 0 | 0 | 0 |
| $S_{14}$ | 0 | 0 | 0 | 1 | 0 | 0 | 0 | 1 |
| $S_{15}$ | 1 | 0 | 1 | 1 | 1 | 0 | 0 | 0 |
| $S_{16}$ | 0 | 1 | 1 | 0 | 1 | 1 | 1 | 0 |

| Population(Query Plan) | |
|---|---|
| | [$R_1$, $R_2$, $R_3$, $R_4$, $R_5$, $R_6$, $R_7$, $R_8$] |
| 1 | [1,1,2,2,2,3,5,3] |
| 2 | [3,5,7,15,4,6,8] |
| 3 | [6,7,8,11,16,6,8,9] |
| 4 | [10,11,11,15,16,11,16,14] |
| 5 | [8,8,10,14,16,11,11,14] |
| 6 | [15,12,13,11,15,15,11,12] |
| 7 | [1,2,5,7,2,4,6,8] |
| 8 | [3,5,7,8,15,3,5,3] |
| 9 | [2,2,2,2,3,5,3] |
| 10 | [2,1,113,2,15,3,5,3] |
| 11 | [8,8,16,14,16,15,16,14] |
| 12 | [3,7,7,8,16,7,16,3] |
| 13 | [2,2,2,2,15,16,14] |
| 14 | [1,1,8,8,2,7,8,8] |
| 15 | [8,8,8,2,7,8,9] |
| 16 | [5,16,15,15,16,6,8,9] |
| 17 | [1,1,1,1,15,15,16,14] |
| 18 | [10,11,11,8,7,3,5,3] |
| 19 | [15,16,15,15,15,16,14] |
| 20 | [1,1,8,8,1,7,8,8] |

For a user query (Q1), as shown below. Multiple alternatives are possible as a relation is stored in multiple sites. The initialization of query equivalents is according to the allocation schema (Relation Site Matrix), as shown in table 1 (a). All the generated query equivalent plans are semantically valid and retrieve the similar results for the user. In table 1 (b) 20 such valid QEPs are listed for the user query (Q1) based on allocation schema given in table 1(a),

**Q1: SELECT** a, m
 **FROM** $R_1$, $R_2$, $R_3$, $R_4$, $R_5$, $R_6$, $R_7$, $R_8$,
 **WHERE** $R_1.a=R_4.t$
 **AND** $R_4.p=R_2.x$ **AND** $R_1.a\_R_7.q$
 **AND** $R_2.x=R_3.n$ **AND** $R_4.x=R_5.s$
 **AND** $R_8.w=R_6.d$ **AND** $R_7.j=R_6.k$

There are two important aspects of a query equivalent plan, namely the content of the query plan and the length/size of the query plan. The size of the query plan is equal to the total number of relations in the user query (in the FROM clause). So, for a user query accessing 4 database relations will have all query equivalents of size 4. Similarly above user query (Q1) required $R_1$,$R_2$, $R_3$, $R_4$, $R_5$, $R_6$, $R_7$ and $R_8$, thus all QEPs is of size 8, as shown in table 1(b). Another important aspect of QEP is the content of QEP. The information in the QEP represents the site name or site id for specific relation. A query plans represent the set of data sites on which relevant data is stored and result will be aggregated. In a query plan [1,1,2,2,2,3,5,3], relations $R_1$ and $R_2$ are accessed from site $S_1$, relations $R_3$, $R_4$ and $R_5$ are accessed from site $S_2$, relations $R_6$ and $R_8$ are accessed from site $S_3$ and relation $R_7$ is accessed from site $S_5$. In a query plans, most significant position site indicate the name of sources site for the $R_1$ and similarly least significant position is for $R_8$.

Next step involved the computation of the associated cost with each design objectives for each QEP. Three design objectives are applied as cost models on generated QEPs. The computation of costs (QAC, QLC, and LPC) values are dependent on the availability of dependent variables required in each of the cost equations. QAC can be computed at compile time, but computation of QLC and LPC is entirely dependent the size of intermediate results of query (selectivity of database operators). In proposed approach, the predicate selectivity is used to estimate the size of intermediate results. Due to which the computation QLC and LPC is also computed at compile time for query optimization. In table 2(a) the 20 QEPs are shown with initial computed cost values, similarly for the remaining QEPs the cost can be computed and further used for the optimization.

## 3. TEACHER LEARNER BASED OPTIMIZATION

Teacher Learner Based Optimization (TLBO) algorithm is a teaching learning process inspired algorithm recently proposed in [20],[21],[22]. It is based on the effect or influence of a teacher on the output of learners or students in a class. The teacher-learner is one of well conventional process of continuous improvement of learner. In the process of learning there are no external parameters required to regulated or provided. These assumptions are the core motivation behind the conception of TLBO as a optimization algorithm. TLBO consider a group of learners as population and different subjects offered to the learners are considered as different design objective. The fitness of learner is having analogy with the marks obtained by the learner on the specific subject. The optimization is done the basis of fitness value. The best solution in the entire population is considered as the teacher. The design objectives are actually the parameters involved in the objective function of the given optimization problem and the best solution is the best value of the objective function. The working of TLBO algorithm is alienated into two phases, 'Teacher phase' and 'Learner phase'. The principle of working of both phases is described in [20],[21], these research paper are conceptually proposed TLBO The multi-objective unconstrained and constrained test functions in this paper, and the results were compared with other optimization algorithms [22].

In the process of optimization, all evolutionary and optimization methods requires optimization related computation. Most of the evolutionary optimization techniques and swarm intelligence-based techniques involved optimization related computation, which is probabilistic. Along with computation these techniques requires controlling or tuning of common or global parameters, such as search space size, number of generations, elite size, etc. In addition to the common control parameters, some algorithm requires regulation of algorithm-specific control-parameters required for better optimize solutions. In genetic algorithm, algorithm- specific parameters such as rate of mutation and crossover rate and mutation rate, similarly, inertia weight and social parameters in particle swarm optimization (PSO). The correct modulation of these parameters is critical for generation of a global optimal solution. The improper regulation of algorithm-specific parameters may lead to increases in computational effort or yields a localized solution. The development of TLBO approaches eliminated the regulation of algorithm-specific control parameters, as there is no algorithm specific parameter in the algorithm [21], [22], [23].So, overall optimization effort is reduced in TLBO and the burden of tuning control parameters is comparatively less in the TLBO algorithm.

**TLBO Algorithm (Optimal Query Plans generation):**

1. Identify criterion(C) or objective functions to solve the given problem-as Query Affinity Cost(QAC),Query Localization(Communication) Cost(QLC) and Local Processing Cost(LPC).
2. Initialize all possible Valid (Semantically equivalent) Query equivalent Plans(QEPs) based on allocation schema or (RSM), where total number of QEP is n.
3. Calculate criterion values vector(C) for all possible QEP.
4. Design matrix (x) with Columns as Number of Criteria and rows as number of QEP.
5. **Teacher Phase**:
   Consider $M_j$ to be the mean, and $T_i$ to be the teacher at any iteration i.
6. Select the best learner as a teacher and calculate mean result of learners in each criteria and $T_i$ will try to improve existing mean $M_j$ towards it so the new mean will be designated as $M_{new}$ and the difference between the existing mean and new mean can be calculated as-
   Difference Mean$_i$ = $r_i$ ($M_{new}$-$T_f$ *$M_j$). Where $T_f$ is as shown,
   $T_f$ =round(1+ r and(0,1){2-1}.
7. Based on this Difference Mean, the existing solution is updated according to the following expression:
   $X_{new,i} = X_{old,i}+\text{Difference\_Mean}_i$
8. **Learner Phase**:
9. Update the learners' knowledge by utilizing the knowledge of some other learner according to equations below, at any iteration *i*, considering two different learners *Xi* and *Xj* where i != j
   $X_{new,i} = X_{old,i}+r_i(X_i-X_j)$ if $f(X_i) < f(X_j)$
   $X_{new,i} = X_{old,i}+r_i(X_i-X_j)$ if $f(X_i) < f(X_j)$
   Accept $X_{new}$ with better function value.
10. Repeat steps 5 to 8 until termination criteria met.

Thus, the TLBO algorithm is simple, effective and involves comparatively less computational effort. TLBO algorithm has been already tested over several constrained and unconstrained benchmark functions for various engineering domain and proved better than the other advanced optimization techniques [23]. It is also proved comparatively better in various field of engineering and their related optimization problems, some are reported in [23],[24].In the field of electrical engineering, [25] in the field of civil engineering. TLBO is primarily designed to handle the optimization problem related to the manufacturing process. Even though [26] raised some doubts about the acceptance of generated solution and the fact that this optimization algorithm does not required tuning or regulation of algorithm-specific parameter. However, in authors had already [22] had already cleared all those issues and justified that the TLBO algorithm does not involved any algorithm-specific parameters. In the literature, it is observed that, the TLBO algorithm is not yet applied in the optimization related problem of database or not used in the generation of 'Optimal Query' plans for distributed query processing. In this paper we have adapted TLBO for the generation of optimal query plans of a user query in the distributed database systems. The initialized query equivalents (QEPs) are analogous to the learners and their cost values are equivalent to performance. In figure 1 the TLBO for optimal query generation is illustrated.

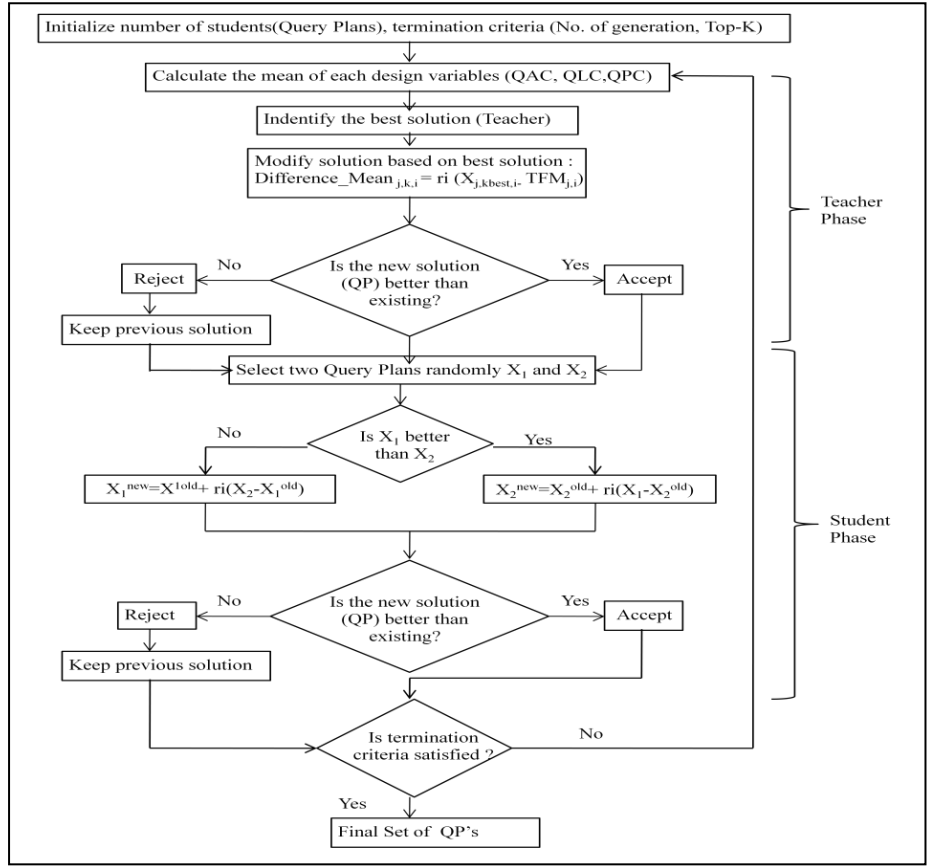

Figure 1. Schematic TLBO for Optimal Query Plan generation in Distributed Database Systems

The inputs to the TLBO are valid QEPs initialized according to the allocation schema and their cost values; few are listed in table 1(a). These query plans (QEPs) in solution space are equivalent to student or learners in TLBO. The TLBO consists of two phases: Teacher phase and Learner phase. In the teacher phase algorithm, students or learners learn via the teacher. A teacher attempts to increase the mean result of the class in the subject he teaches, depending on his or her capability. Different design objective are analogue is to the different subjects taught by teacher and the respective subject values represent the score of particular student or learner in the subject. The best overall result according to the all the subjects together obtained in the entire population of learners can be considered the result of the best learner [21]. For the calculation of final fitness a benchmark function is used, in our implementation we have employed. However, since the teacher is usually considered a highly learned person who trains learners so that they can have better results, the algorithm considers the best identified learner as teacher [22]. This best learner tries to improve the mean results of the entire set of learners in the subsequent phase of algorithm. The identification of best learner's based on the final fitness value of learner, fitness value is evaluated by benchmark function. In our problem learner with minimum fitness values considered best in set, as in query optimization it is desired to select solution with lesser cost values. So query plan no. 15 is best among learners, shown in table 2 (d). Now difference mean values or teacher factor ($T_F$) are to be calculated, which decides the degree of change on each students or learner. Teaching factor ($T_f$) is the difference between the existing mean results of each subject and the corresponding result of the teacher for each subject, as shown in table 2 (c) and values of ($T_F$) either 1 or 2, as $T_F$ = round [1+rand(0,1),{2-1}]. Based on the calculated $T_F$, new population is evolved. The values of each the subject values (objective function) are modified according to the computed $T_F$ on each criterion's. New population, as shown in table 2 (e) go through learner phase.

Next phase is student phase, in which students or learners raise their knowledge level by interaction among themselves. A learner or student interacts randomly with other learners to enhance his or her knowledge. Core philosophy behind this phase is that a learner learns new things if the other learner has more knowledge than him or her. The updated population (table 2 (e)) is the input to this phase, algorithm randomly selects pair of the students and compare based on the fitness value ($fitness_2$), winner of the comparison ($QP_1$ better than $QP_2$, or vice versa) will update the weaker student's cost values in pair. The students with better subject's values will be improving the subjects values of weaker students, similarly each student go through this step at least once in algorithm run. This is purely summarizing the mutual learning in the class room learning environment.

Table 2
Teacher-Phase related results: (a) Initial Population and Costs (b) Mean values of Design Objectives (c) Teaching Factor or Difference Factor ($T_F$) (d) Best Teacher Query Plan (e) Updated population based on $T_F$

| Teacher Phase |||||||||||
|---|---|---|---|---|---|---|---|---|---|---|
| Initial Population(Query Plan) || Costs ||| Final Fitness$_1$ | Mean Value || QEP No. | Updated Population_01 |||
| QEP No. | Query Plan | QAC | QLC | LPC | | QAC | 0.7015 | | QAC | QLC | LPC |
| 1 | [1,1,2,2,2,3,5,3] | 0.7188 | 0.5354 | 0.2667 | 0.8744 | QLC | 0.4797 | 1 | 0.1281 | -0.2506 | 0.0868 |
| 2 | [3,5,7,15,4,6,8] | 0.8438 | 0.6691 | 0.3466 | 1.2797 | LPC | 0.2999 | 2 | 0.2531 | -0.1169 | 0.1667 |
| 3 | [6,7,8,11,16,6,8,9] | 0.8125 | 0.5374 | 0.3878 | 1.0994 | | | 3 | 0.2219 | -0.2486 | 0.2079 |
| 4 | [10,11,11,15,16,11,16,14] | 0.7500 | 0.4315 | 0.3853 | 0.8971 | Diff. Factor($T_F$) || 4 | 0.1594 | -0.3545 | 0.2054 |
| 5 | [8,8,10,14,16,11,11,14] | 0.7813 | 0.6691 | 0.3724 | 1.1967 | QAC | -0.5906 | 5 | 0.1906 | -0.1169 | 0.1925 |
| 6 | [15,12,13,11,15,15,11,12] | 0.7188 | 0.6691 | 0.2615 | 1.0327 | QLC | -0.7860 | 6 | 0.1281 | -0.1169 | 0.0816 |
| 7 | [1,2,5,7,2,4,6,8] | 0.8438 | 0.6691 | 0.4651 | 1.3760 | LPC | -0.1799 | 7 | 0.2531 | -0.1169 | 0.2853 |
| 8 | [3,5,7,8,15,3,5,3] | 0.7500 | 0.6691 | 0.3337 | 1.1215 | | | 8 | 0.1594 | -0.1169 | 0.1538 |
| 9 | [2,2,2,2,3,5,3] | 0.5313 | 0.3130 | 0.1430 | 0.4007 | Best Query Plan || 9 | -0.0594 | -0.4730 | -0.0369 |
| 10 | [2,1,113,2,15,3,5,3] | 0.8125 | 0.6691 | 0.3517 | 1.2316 | 15 || 10 | 0.2219 | -0.1169 | 0.1719 |
| 11 | [8,8,16,14,16,15,16,14] | 0.7188 | 0.5215 | 0.2899 | 0.8726 | | | 11 | 0.1281 | -0.2645 | 0.1100 |
| 12 | [3,7,7,8,16,7,16,3] | 0.7188 | 0.4315 | 0.3054 | 0.7960 | | | 12 | 0.1281 | -0.3545 | 0.1255 |
| 13 | [2,2,2,2,15,16,14] | 0.5625 | 0.3130 | 0.1945 | 0.4522 | | | 13 | -0.0281 | -0.4730 | 0.0146 |
| 14 | [1,1,8,8,2,7,8,8] | 0.6563 | 0.3091 | 0.2925 | 0.6117 | | | 14 | 0.0656 | -0.4770 | 0.1126 |
| **15** | **[8,8,8,2,7,8,9]** | **0.4063** | **0.0867** | **0.2100** | **0.2167** | | | 15 | -0.1844 | -0.6993 | 0.0301 |
| 16 | [5,16,15,15,16,6,8,9] | 0.8125 | 0.5513 | 0.4007 | 1.1247 | | | 16 | 0.2219 | -0.2348 | 0.2208 |
| 17 | [1,1,1,1,15,15,16,14] | 0.6563 | 0.3289 | 0.2667 | 0.6100 | | | 17 | 0.0656 | -0.4571 | 0.0868 |
| 18 | [10,11,11,8,7,3,5,3] | 0.8125 | 0.5023 | 0.4136 | 1.0835 | | | 18 | 0.2219 | -0.2837 | 0.2337 |
| 19 | [15,16,15,15,15,16,14] | 0.5313 | 0.4090 | 0.1430 | 0.4700 | | | 19 | -0.0594 | -0.3770 | -0.0369 |
| 20 | [1,1,8,8,1,7,8,8] | 0.5938 | 0.3091 | 0.1688 | 0.4765 | | | 20 | 0.0031 | -0.4770 | -0.0111 |

The outcome of student phase is a new population of QEPs with updated cost values, as shown in table 3(b). This completes the first iteration of TLBO. After a number of sequential run of teaching–learning cycles in which, teacher disseminates knowledge to the learners and their knowledge level increases toward the teacher's level. The distribution of the randomness within the search space becomes more and smaller around a point that is considered the teacher. Therefore, the knowledge level of the entire class is smooth and the algorithm converges towards an optimal solution or pareto set. The query plans cost values are normalized and updated in subsequent run of TLBO and finally reaches to the point of optimality in the solution space. In the final step, TLBO terminates according to the provided values of termination criteria's same any other optimization technique. In the query optimization, the termination criteria's such as Top-K query plans, Query Cost, No. of Generation etc values are supplied to TLBO for termination. The final ranks of QEPs, symbolize the fitness measure of specific QEP for result generation for user query, as the cost values for the QEP optimal, as shown in table 3 (c). The QEPs are ranked based on the QC value; QC values are computed using the similar benchmark function as for the initial fitness function. Further, selected QEP or pareto front are used for query result generation and thus supplied to query executor/processor. The metadata related to the sites involved are fetched and supplied to query compiler and executor. The query processor retrieves the sub results and integrates to a control site for the generation of user query result. The final result of user query is arrange in the structure user requested and send back to the user or application. The query plans is kept in directory for future reference. The query processor usually keeps records of the user query for the performance improving or optimization in the future. In the paper we compared the performance of GA based solution for the optimal plans generation with the TLBO based heuristic. TLBO is proven better than GA based solution.

Table 3

Student-Phase related results: (a) Population from table 2.(e) with Cost values (b) updated Population_02 based on the Mutual Learning (c)Ranked Query Plans (among 20 QEPs) after 1st generation of TLBO

| QEP No. | Updated_Population_01(Cost-Matrix) | | | Fitness$_2$ | QEP No. | Updated_Population_02() | | | Ranked Query Plans | | |
|---|---|---|---|---|---|---|---|---|---|---|---|
| | QAC | QLC | LPC | | | QAC | QLC | LPC | QEP No. | Rank | QC Cost |
| 1 | 0.1281 | -0.2506 | 0.0868 | 0.0868 | 1 | 0.1281 | -0.2266 | 0.1075 | 15 | 1 | 0.0918 |
| 2 | 0.2531 | -0.1169 | 0.1667 | 0.1055 | 2 | 0.2463 | -0.2098 | 0.1667 | 19 | 2 | 0.1325 |
| 3 | 0.2219 | -0.2486 | 0.2079 | 0.1543 | 3 | 0.2219 | -0.2486 | 0.0970 | 9 | 3 | 0.1474 |
| 4 | 0.1594 | -0.3545 | 0.2054 | 0.1933 | 4 | 0.1594 | -0.3736 | 0.2054 | 13 | 4 | 0.1721 |
| 5 | 0.1906 | -0.1169 | 0.1925 | 0.0871 | 5 | 0.1730 | -0.1169 | 0.1925 | 20 | 5 | 0.2909 |
| 6 | 0.1281 | -0.1169 | 0.0816 | 0.0368 | 6 | 0.1281 | -0.1169 | 0.0816 | 17 | 6 | 0.3375 |
| 7 | 0.2531 | -0.1169 | 0.2853 | 0.1591 | 7 | 0.2531 | -0.3327 | 0.0922 | 11 | 7 | 0.3434 |
| 8 | 0.1594 | -0.1169 | 0.1538 | 0.0627 | 8 | 0.1594 | -0.1169 | 0.1538 | 14 | 8 | 0.3733 |
| 9 | -0.0594 | -0.4730 | -0.0369 | 0.2286 | 9 | -0.0505 | -0.4552 | 0.0207 | 6 | 9 | 0.3736 |
| 10 | 0.2219 | -0.1169 | 0.1719 | 0.0924 | 10 | 0.2219 | -0.1169 | 0.1719 | 12 | 10 | 0.3794 |
| 11 | 0.1281 | -0.2645 | 0.1100 | 0.0985 | 11 | 0.1536 | -0.2645 | 0.1100 | 1 | 11 | 0.3981 |
| 12 | 0.1281 | -0.3545 | 0.1255 | 0.1579 | 12 | 0.2036 | -0.2883 | 0.0827 | 4 | 12 | 0.4447 |
| 13 | -0.0281 | -0.4730 | 0.0146 | 0.2247 | 13 | 0.1677 | -0.3446 | 0.0146 | 8 | 13 | 0.5661 |
| 14 | 0.0656 | -0.4770 | 0.1126 | 0.2445 | 14 | 0.0739 | -0.4319 | 0.1261 | 5 | 14 | 0.5788 |
| 15 | -0.1844 | -0.6993 | 0.0301 | 0.5240 | 15 | -0.0894 | -0.2835 | 0.1184 | 16 | 15 | 0.6193 |
| 16 | 0.2219 | -0.2348 | 0.2208 | 0.1531 | 16 | 0.2219 | -0.2768 | 0.2208 | 3 | 16 | 0.6278 |
| 17 | 0.0656 | -0.4571 | 0.0868 | 0.2208 | 17 | 0.0735 | -0.4571 | 0.1367 | 18 | 17 | 0.6676 |
| 18 | 0.2219 | -0.2837 | 0.2337 | 0.1843 | 18 | 0.2080 | -0.2837 | 0.1562 | 7 | 18 | 0.6959 |
| 19 | -0.0594 | -0.3770 | -0.0369 | 0.1470 | 19 | -0.0594 | -0.3178 | -0.0369 | 10 | 19 | 0.6964 |
| 20 | 0.0031 | -0.4770 | -0.0111 | 0.2276 | 20 | 0.0699 | -0.4770 | -0.0111 | 2 | 20 | 0.7355 |

## 4. RESULT ANALYSIS

The experimental analysis for performance comparison is based on the three approaches. The TLBO based solution and two GA based solution are compared. GA based multi- objective approach, such as aggregation and vector evaluated genetic algorithm (VEGA) are adapted. As mentioned earlier also, in GA based solution algorithm-specific parameter such as crossover and mutation probability are required to be regulated during the optimization while in TLBO based solution no such parameter are to regulated. In below graphs various criteria are used to illustrate the performance for optimal query plans generation.

*4.1. Top-K Query Plans generation*

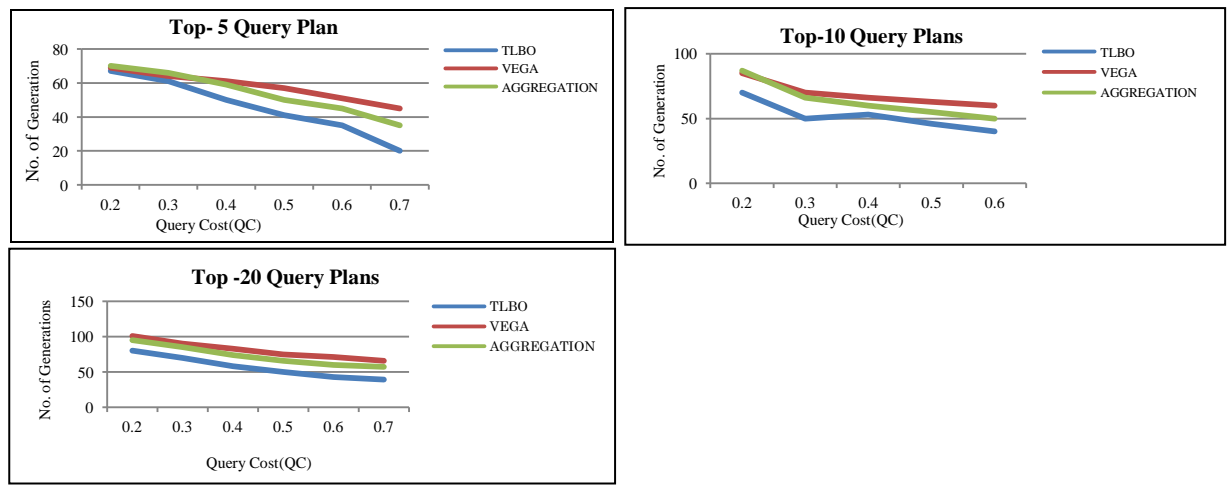

Figure 2. Generation of Top-K (K=5,10, 20) QEPs for Query Cost (QC= 0.2,0.3,0.4,0.5, 0.6.0.7) with no. of evolution or iterations on Algorithms (TLBO, Multi-Objective VEGA and Aggregation based GA with, crossover probability ($P_c$)= 0.8, mutation probability ($P_m$)=0.2, Weight$_{QAC}$=0.2, Weight$_{QLC}$=0.5, Weight$_{LPC}$=0.3)

*4.2. for a given Query Cost of 'Optimal Query' plan generation*

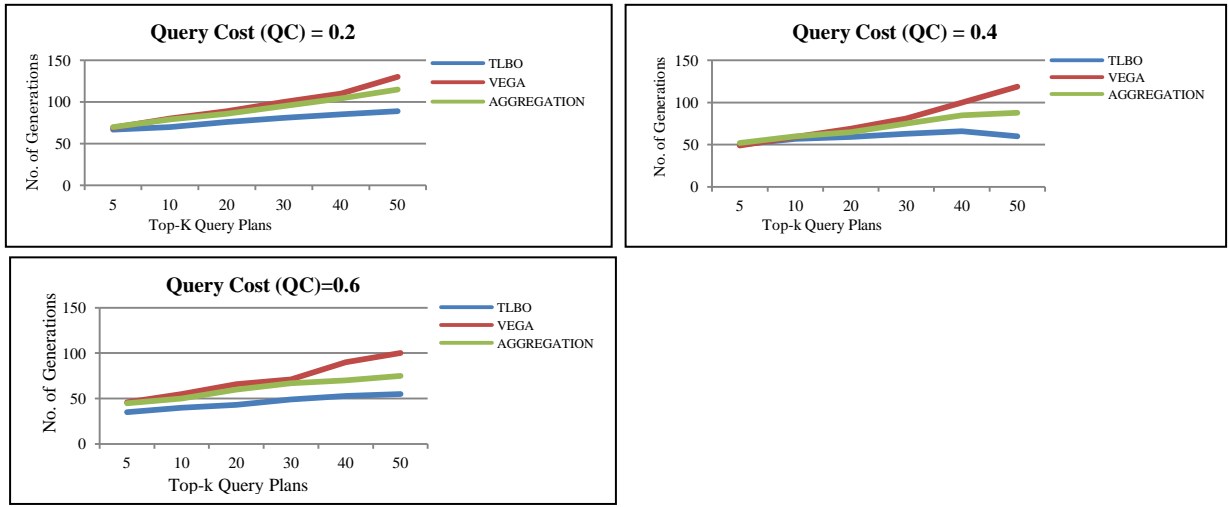

Figure 3. For given Query Cost (QC= 0.2, 0.4, 0.6) generation of Top-K (K=5, 10, 20, 30, 40, 50) Query plans for in No. of iterations or generations of Algorithm (TLBO, Multi-Objective VEGA and Aggregation based GA, crossover probability ($P_c$)= 0.8, mutation probability ($P_m$)=0.2, Weight$_{QAC}$=0.2, Weight$_{QLC}$=0.5, Weight$_{LPC}$=0.3)

*4.3. Generation of 'Optimal Top-K' in different number of sites ($N_s$) and relations ($N_r$)*

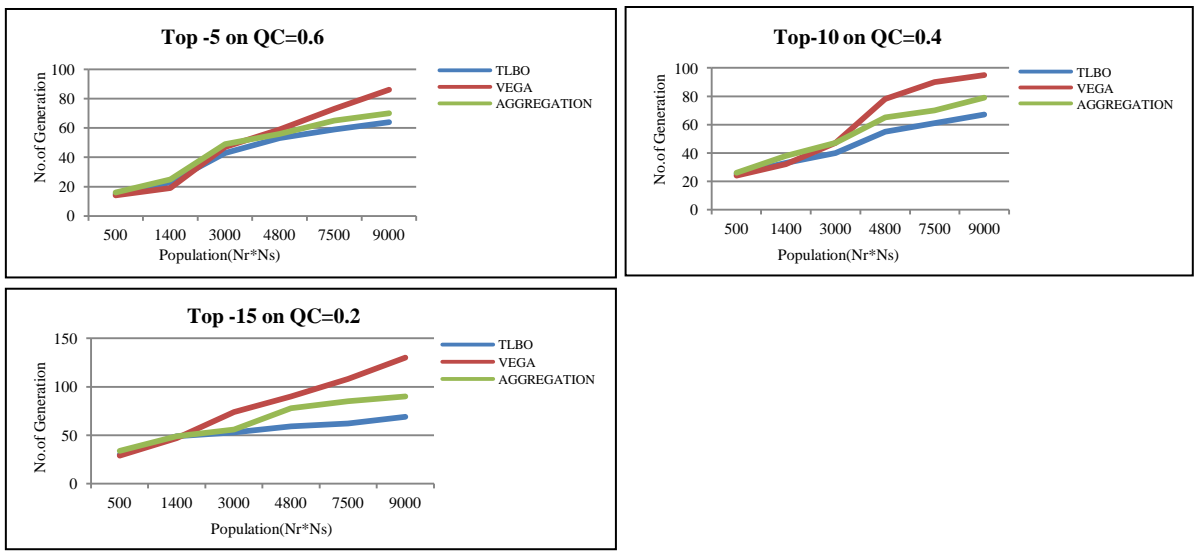

Figure 4. Generation of 'Optimal Top-K' query plans with Query Cost (K=5 & QC= 0.6, K=10 & QC= 0.4, K=15& 0.2) for different Number of Sites ($N_s$) and relations sizes ($N_r$) in number of iterations or generations of Algorithm (TLBO, Multi-Objective VEGA and Aggregation based GA with, crossover probability ($P_c$)= 0.8, mutation probability ($P_m$)=0.2, Weight$_{QAC}$=0.2, Weight$_{QLC}$=0.5, Weight$_{LPC}$=0.3)

## 6. CONCLUSIONS

In distributed database system data is dispersed over the multiple sites, this distribution of data is based on partition or replication based due to which a given relation can be found in more than one sites. Query processing in such environment is difficult task for query processor, as multiple equivalent alternatives. Query processing in such environment, major objective design are CPU, I/O and the site-to-site communication cost, among these, the site-to-site communication cost is the dominant cost. This requires an optimization mechanism to generate optimal set of query plans to retrieve results for user query. Multi-objective optimization is a very important research area in engineering studies, because real-world design problems require the optimization of a group of objectives. Multiple, often conflicting, objectives arise naturally in most real-world optimization scenarios. Adding more than one objective to an optimization problem adds complexity. In this paper, TLBO algorithm is employed to generate optimal top-k query plans for proposed design objectives and in unconstrained function and its performance is compared with nature-inspired optimization technique, genetic algorithm (GA). The experimental results show that the TLBO performs competitively better on the generation top-k query plans with other optimization methods. Therefore, the TLBO algorithm is effective and robust and has a great potential for solving similar multi-objective problems. The optimality on generation of query plans by other swarm based optimization techniques is part of our future work based on the similar design objectives.

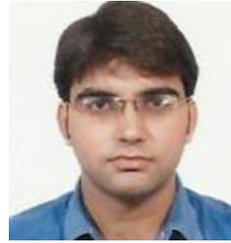
Vikash Mishra has completed master's degree in Computer Engineering with specialization of Database Systems and applications from National Institute of Technology Kurukshetra, Haryana, India in, June 2015. He is currently working as Junior Associate Software Developer in Nagarro Software Pvt. Ltd., Gurgaon, Haryana, India. His areas of interest are mostly on Database Systems, Big Data Analytics, and Optimization techniques.

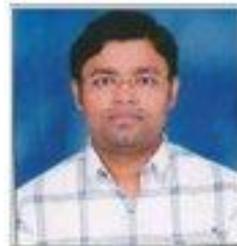
Mr. Vikram Singh is Assistant Professor in the Computer Engineering Department, National Institute of Technology, Kurukshetra, Haryana, India since year 2012.

His research interests are in Advanced Database Systems, Big Data Analytics, Query Processing and Optimization, DataSapce systems, Personal Information management Systesm. He is a Member of the IEEE and ACM Society.